\documentclass[10pt,conference]{IEEEtran}
\usepackage{epsfig,setspace,amsmath,epsf,amssymb,bm,theorem,cite, graphicx, epstopdf, algorithm, algpseudocode,float,color,physics,cuted,comment}
\usepackage{multirow}
\usepackage{authblk}
\usepackage[table,xcdraw]{xcolor}
\usepackage{mathtools}

\newtheorem{theorem}{Theorem}

\newtheorem{definition}{Definition}
\newtheorem{remark}{Remark}

\IEEEoverridecommandlockouts

\newcommand{\cw}{{\mathcal{W}}}
\newcommand{\cs}{{\mathcal{S}}}

\DeclareMathOperator{\diag}{diag}

\allowdisplaybreaks

\begin{document}

\title{Quantum Symmetric Private Information Retrieval with Secure Storage and Eavesdroppers}
\author{Alptug Aytekin \quad Mohamed Nomeir \quad Sajani Vithana \quad Sennur Ulukus\\
	\normalsize Department of Electrical and Computer Engineering\\
	\normalsize University of Maryland, College Park, MD 20742 \\
	\normalsize \emph{aaytekin@umd.edu} \quad \emph{mnomeir@umd.edu} \quad \emph{spallego@umd.edu} \quad \emph{ulukus@umd.edu}}

\maketitle

\begin{abstract}
  We consider both the classical and quantum variations of $X$-secure, $E$-eavesdropped and $T$-colluding symmetric private information retrieval (SPIR). This is the first work to study SPIR with $X$-security in classical or quantum variations. We first develop a scheme for classical $X$-secure, $E$-eavesdropped and $T$-colluding SPIR (XSETSPIR) based on a modified version of cross subspace alignment (CSA), which achieves a rate of $R= 1 - \frac{X+\max(T,E)}{N}$. The modified scheme achieves the same rate as the scheme used for $X$-secure PIR with the extra benefit of symmetric privacy. Next, we extend this scheme to its quantum counterpart based on the $N$-sum box abstraction. This is the first work to consider the presence of eavesdroppers in quantum private information retrieval (QPIR). In the quantum variation, the eavesdroppers have better access to information over the quantum channel compared to the classical channel due to the over-the-air decodability. To that end, we develop another scheme specialized to combat eavesdroppers over quantum channels. The scheme proposed for $X$-secure, $E$-eavesdropped and $T$-colluding quantum SPIR (XSETQSPIR) in this work maintains the super-dense coding gain from the shared entanglement between the databases, i.e., achieves a rate of $R_Q = \min\left\{ 1, 2\left(1-\frac{X+\max(T,E)}{N}\right)\right\}$.
\end{abstract}

\section{Introduction}

In the private information retrieval (PIR) problem introduced in \cite{chor}, a user wishes to retrieve a message out of $K$ messages stored in $N$ databases without revealing the index of the required message to any of the databases. The optimal rate of PIR with $N$ databases and $K$ replicated messages is shown to be $C(N,K) = (1+\frac{1}{N}+\ldots+\frac{1}{N^{K-1}})^{-1}$ in \cite{c_pir}. Subsequently, several variations of this problem have been studied with different requirements for the databases and the user. In \cite{c_spir}, symmetric PIR (SPIR) is introduced, where the user is not allowed to obtain any information about the message set other than the required message. The capacity of SPIR is shown to be $1-\frac{1}{N}$ in \cite{c_spir}, which is also $C(N,\infty)$. In \cite{mdstpir}, $T$-colluding PIR is introduced where any $T$ databases can share the queries received from the user to learn the required message index. The capacity of $T$-colluding PIR is shown to be $(1+\frac{T}{N}+\ldots+\frac{T^{K-1}}{N^{K-1}})^{-1}$ in \cite{mdstpir}, which is also $C(\frac{N}{T},K)$. $T$-colluding SPIR is considered in \cite{tspir_mdscoded} and its capacity is shown to be $1-\frac{T}{N}$, which is $C(\frac{N}{T},\infty)$. In \cite{C_SETPIR}, the $E$-eavesdropped, $T$-colluding SPIR is introduced. In this setting, there is an eavesdropper that can listen to all answers from any $E$ databases to the user. The capacity for this case is shown to be $1-\frac{\max(T,E)}{N}$ in \cite{C_SETPIR}. The problem of $X$-secure PIR is introduced in \cite{first_xsecure}, where the messages need to be hidden from the databases themselves even when $X$ databases share their complete datasets. In \cite{csa}, the asymptotic capacity of $X$-secure $T$-colluding PIR, i.e., the capacity when  $K\rightarrow\infty$, is shown to be $1-\frac{X+T}{N}$. Some other variations of the PIR and SPIR problems have been studied and different applications have been introduced in \cite{arbitrarycollusion, byzantine_tpir, banawan_eaves, banawan_multimessage_pir, banawan_pir_mdscoded, uncoded_constrainedstorage_pir, batuhan_hetero,  grpahbased_pir, ITW_paper, multimessage_pir_sideinfo, wei_banawan_cache_pir, wang_spir, semantic_pir,salim_singleserver_pir,Salim_CodedPIR}; see also \cite{recent}. 

The problem of quantum PIR (QPIR) is recently introduced in \cite{qpir}. In this model, the message bits are sent over a quantum channel from the databases to the user, and the databases can share entanglement between them. \cite{qpir} shows that the capacity of symmetric QPIR (SQPIR) is $1$ when the number of databases is $N\geq 2$. Variations of QPIR include $T$-colluding QPIR with and without coded storage \cite{qtpir, qtpir_t=n-1, qpir_colluding_mdscoded}, QPIR with noisy channels \cite{noisy_qpir}, and several other variations analogous to their classical counterparts \cite{qmqpir, qpir_byzantine_colluding, qpir_star_product_codes}. Most recently, \cite{nsumbox} has proposed a mathematical abstraction for the entanglement between transmitters sending information to a common receiver over separate quantum channels. The work in \cite{nsumbox} shows that the entanglement between $N$ transmitters that use Pauli operators to encode classical messages to quantum states can be represented mathematically as a multiple input multiple output (MIMO) multiple access channel (MAC) with $2N$ inputs and $N$ outputs, i.e., a matrix with  $N\times 2N$ dimensions. In addition, this matrix must have elements from a finite field, and must satisfy the strong self orthogonal (SSO) property. Using these properties, \cite{nsumbox} shows that the rate of $X$-secure $T$-colluding QPIR for their proposed scheme is $R_Q = \min\left\{1,2\left(1-\frac{X+T}{N}\right)\right\}$. This is a doubling of the classical rate $R_C=1-\frac{X+T}{N}$ in the regime of interest.

In this paper, we focus on both classical and quantum variations of the SPIR problem with a passive eavesdropper which listens to queries and answers going into and out of any of the $E$ databases. In addition, up to $T$ databases collude, and up to $X$ databases communicate. This is the first work that considers $X$-secure SPIR in general, even in the classical domain, and even without $E$-eavesdropped and $T$-colluding databases. We show that the rate of the modified CSA scheme (modified for the symmetric privacy) $R$ is the same as the rate of the CSA scheme proposed in \cite{csa} with the extra benefit of symmetric privacy, i.e., $R= 1 - \frac{X+\max(T,E)}{N}$. In addition, this is the first work to consider the presence of eavesdroppers in QPIR. We develop a QPIR scheme that maintains privacy and security against eavesdroppers, colluding databases and communicating databases. Our proposed quantum scheme achieves the maximum super-dense coding gain when an entangled state is shared between the databases, i.e., $R_Q = \min\left\{ 1, 2\left(1-\frac{X+\max(T,E)}{N}\right)\right\}$. The QPIR problem with eavesdroppers is more complex compared to the classical PIR problem with eavesdroppers due to the over-the-air decodability imposed by the $N$-sum box abstraction. The quantum scheme we propose in this work achieves double the rate of its classical counterpart.

\section{Preliminaries}
In this section, we state some important definitions related to quantum physics and quantum information theory \cite{nielsen-chuang}. We use these quantities subsequently to formulate the $X$-secure $E$-eavesdropped $T$-colluding QSPIR.

\begin{definition}[Quantum density matrices]
    For a general quantum system $A$, that can be in the state $\ket{\psi_j}$ with probability $p_j$, the quantum density matrix $\rho_A$ is defined as,
\begin{align}
    \rho_A = \sum_j p_j \ket{\psi_j}\bra{\psi_j},
\end{align}
with $p_j \geq 0$, $\sum_j p_j =1$.
\end{definition}

\begin{definition}[Von Neumann entropy]
    For the density matrix $\rho$, Von Neumann entropy is defined as,
    \begin{align}
        S(\rho)= -tr(\rho \log\rho)=H(\Lambda),
    \end{align}
    where $tr(.)$ is the trace operator, $\Lambda$ are the eigenvalues of $\rho$, and $H(\cdot)$ is the Shannon entropy. For a quantum system $A$ with density matrix $\rho_A$, we define $S(A) = S(\rho_A)$.
\end{definition}

\begin{definition}[Quantum relative entropy]
    The relative entropy between two density matrices $\rho$ and $\sigma$ is defined as, 
    \begin{align}
        D(\rho \lVert \sigma) = tr(\rho (\log\rho-\log\sigma)).
    \end{align} 
\end{definition}

\begin{definition}[Quantum conditional entropy]
    The conditio-nal entropy of a quantum system $A$ with respect to a system $B$ is defined as,
    \begin{align}
        S(A|B) = S(A,B)-S(B).
    \end{align}
    \end{definition}
\begin{definition}[Quantum mutual information]
    The quantum mutual information between two quantum systems $A$ and $B$ is defined as,
    \begin{align}
        S(A;B)&=S(A)+S(B)-S(A,B)\\
        &=S(A)-S(A|B).
    \end{align}
\end{definition}

In the next section, we formulate the problem in both classical and quantum variations.

\section{Problem Formulation}

The system consists of $N$ databases and a user who wants to retrieve a message. Out of the $N$ databases, $T$ are allowed to collude, i.e., share the user's queries, and $X$ are allowed to communicate, i.e., share their storage to decode the messages. In addition, any $E$ links are accessible to the eavesdroppers that can listen to $E$ of the user's queries and databases' answers. The system contains $K$ messages, $W_1,\ldots,W_K$, of equal length $L$, that are independent and identically distributed (i.i.d.). The messages are generated uniformly at random from the field $\mathbb{F}_q$, with $q=p^r$, where $p$ is any prime number. Thus,
\begin{align} 
    H(W_k)&=L, \quad k \in [1:K], \label{message_entropy}\\
    H(W_{[1:K]})&= \sum_{k=1}^{K} H(W_k) = KL. \label{messages_entropy}
\end{align}
The messages $W_{[1:K]}$ need to be secure against any $X$ communicating databases,
\begin{align} \label{x_storage_constraint}
    I(W_{[1:K]}; S_{\mathcal{X}})= 0,
\end{align}
where $S_{\mathcal{X}}$ denotes all of the stored data in subset $\mathcal{X}$ databases satisfying $|\mathcal{X}|\leq X$. The user wants to retrieve a message $W_{\theta}$, where $\theta$ is chosen uniformly at random from $[1:K]$, and sends a query to each database $(Q_{1}^{[\theta]},\ldots,Q_{N}^{[\theta]})$ denoted by $Q_{[1:N]}^{[\theta]}$. As the user does not know the messages, the queries are generated independent of the message content, 
\begin{align}\label{colluding_constraint}
    I(W_{[1:K]}; Q_{[1:N]}^{[\theta]}) = 0, \quad \theta \in [1:K].
\end{align}
In addition, we require that the index of the retrieved message by the user is private against any $T$ colluding databases, 
\begin{align}\label{privacy_constraint}
    I(\theta; Q_{\mathcal{T}}^{[\theta]}) = 0, ~ \theta \in [1:K],
\end{align}
where $\mathcal{T} \subset [1:N], ~ |\mathcal{T}| \leq T$.

Upon receiving the queries, the $n$th database replies with a deterministic answer string $A_{n}^{[\theta]}$ based on its received query $Q_{n}^{[\theta]}$, shared common randomness between the databases $\cs$, and stored data, $S_{n}, ~ n \in [1:N]$, 
\begin{align}\label{classical_answer_encoding} 
    H(A_{n}^{[\theta]}|S_{n},Q_{n}^{[\theta]},\cs) = 0, \quad \theta \in [1:K].
\end{align}
When the user receives all answer strings $A_{[1:N]}^{[\theta]}$, the required message must be decodable based on the answer strings and the sent queries, 
\begin{align}\label{classical_decodaability}
    H(W_{\theta}|A_{[1:N]}^{[\theta]},Q_{[1:N]}^{[\theta]})=0, ~ \theta \in [1:K].
\end{align}
In addition, the symmetric privacy constraint requires that the user gains no information about the message set except for the required message, 
\begin{align}\label{symmetric_privacy}
I(\cw_{\theta^C};A^{[\theta]}_{[1:N]}|Q_{[1:N]}^{[\theta]},\theta) = 0, ~ \theta \in [1:K],
\end{align}
where $\cw_{\theta^C}$ denotes all other messages aside from the required message $W_{\theta}$.

Finally, the scheme must be private and secure against an eavesdropper who can listen to any set of $E$ queries and $E$ answers, 
\begin{align}\label{classical_eavesdropper_uplink}
    &I(\theta;Q^{[\theta]}_{\mathcal{E}_1},A^{[\theta]}_{\mathcal{E}_2}) = 0,\quad \theta \in [1:K],
    \end{align}
    and
    \begin{align}\label{classical_eavesdropper_downlink}
&I(W_{[1:K]};A^{[\theta]}_{\mathcal{E}_1}|Q^{[\theta]}_{\mathcal{E}_2})=0, \quad \theta \in [1:K],
\end{align}
where $\mathcal{E}_1, \mathcal{E}_2 \subset [1:N], ~|\mathcal{E}_1|,|\mathcal{E}_2| \leq E$. 

The rate $R$ of any scheme satisfying the above requirements is defined as the ratio between the length of the required message and the average length of the answer strings,
\begin{align} \label{classical_rate}
    R= \frac{L}{H(A_{[1:N]}^{[\theta]})}.
\end{align}

In the $X$-secure, $E$-eavesdropped, $T$-colluding quantum symmetric PIR (XSETQSPIR) problem, we follow the system models introduced in the literature \cite{qpir, qpir_byzantine_colluding, qtpir,qtpir_t=n-1}. The databases store $S_{n}$, $n \in [1:N]$, as classical bits and share an entangled state of $N$ quantum bits denoted by $\rho$. The user sends the queries $Q_{[1:N]}^{[\theta]}$ over a classical channel to each of the $N$ databases, and each database $n$, $n\in[1:N]$, with the quantum system $\mathcal{A}_n^0 = tr_{j=[1:N]\atop j \neq n}(\rho)$, where $tr(\cdot)$ is the trace operator, replies to the user queries over a separate quantum channel. Upon receiving the query, the $n$th database performs the quantum operation $Enc_n$ based on the received query, storage and $\mathcal{A}_n^0$ to produce the quantum state $\mathcal{A}_n^{[\theta]}$, $n \in [1:N]$, as follows, 
\begin{align}
    \mathcal{A}_n^{[\theta]} = Enc_n(Q^{[\theta]}_n,S_n, \mathcal{A}_n^0, \Lambda_n,\cs), ~ \theta \in [1:K],  
\end{align}
where $Enc_n$ is the $n$th database's encoder, and $\Lambda_n$ is a masking random variable sent by the user to the databases.\footnote{The main reason for the masking random variables is to fight over-the-air decodability in quantum channels. It is discussed in detail in Section~\ref{quantum scheme}.} The final received state at the user is given as, 
\begin{align}
    \mathcal{A}_{[1:N]}^{[\theta]} = \mathcal{A}_1^{[\theta]} \otimes \ldots \otimes \mathcal{A}_N^{[\theta]}, ~ \theta \in [1:K],
\end{align}
where $\otimes$ is the tensor product. Since the storage is in the form of classical bits and the queries are sent over classical channels, constraints \eqref{x_storage_constraint}-\eqref{privacy_constraint} must hold. It is also required that the index of the required message be secure against the received queries and masking random variables $\Lambda_{[1:N]}$ for any $\mathcal{T} \subset [1:N],~ |\mathcal{T}|\leq T$ colluding databases, 
\begin{align}\label{colluding_with_masking}
    I(\theta;Q^{[\theta]}_{\mathcal{T}},\Lambda_{\mathcal{T}}) = 0, ~  \theta \in [1:K].
\end{align}
Additionally, the Von Neumann entropy of the required message $W_{\theta}$ given the queries and the answers must be zero,
\begin{align}\label{quantum_decodability}
    S(W_{\theta}| \mathcal{A}_{[1:N]}^{[\theta]},Q_{[1:N]}^{[\theta]},\Lambda_{[1:N]})=0, ~\theta \in [1:K],
\end{align}
and for symmetric privacy, the quantum mutual information between the other messages $\cw_{\theta^C}$ and the received quantum densities $\mathcal{A}^{[\theta]}_{[1:N]}$ must satisfy,
\begin{align}\label{quantum_symmetric_privacy}
S(\cw_{\theta^C};\mathcal{A}^{[\theta]}_{[1:N]}|Q_{[1:N]}^{[\theta]},\theta,\Lambda_{[1:N]}) = 0, ~ \theta \in [1:K].
\end{align}
In addition, for the eavesdroppers who listen to any $E$ classical and quantum channels, the privacy and security requirements must be satisfied, 
\begin{align}
    S(\theta;Q^{[\theta]}_{\mathcal{E}_1},\mathcal{A}^{[\theta]}_{\mathcal{E}_2}, \Lambda_{\mathcal{E}_3}) &= 0,~ \theta \in [1:K],\label{quantum_eavesdropper_uplink}\\
S(W_{[1:K]};\mathcal{A}^{[\theta]}_{\mathcal{E}_1}|Q^{[\theta]}_{\mathcal{E}_2},\Lambda_{\mathcal{E}_3})&=0,  ~\theta \in [1:K],\label{quantum_eavesdropper_downlink}
\end{align}
where $\mathcal{E}_1,\mathcal{E}_2,\mathcal{E}_3 \subset [1:N]$ and $|\mathcal{E}_1|,|\mathcal{E}_2|,|\mathcal{E}_3| \leq E$. Then, the XSETQSPIR rate $R_Q$ for the retrieval scheme satisfying \eqref{x_storage_constraint}-\eqref{privacy_constraint} and  \eqref{colluding_with_masking}-\eqref{quantum_eavesdropper_downlink} is defined as
\begin{align}
    R_Q = \frac{H(W_{\theta})}{dim(\mathcal{A}_1^{[\theta]}\otimes\ldots\otimes\mathcal{A}_N^{[\theta]})}.
\end{align}

In this paper, we follow the encoding and decoding structure using the $N$-sum box abstraction introduced recently in \cite{nsumbox}. In the encoding stage, the databases use Pauli operators $\mathsf{X}(a) = \sum_{j=0}^{q-1} \ket{j+a}\bra{j}$, and  $\mathsf{Z}(a) = \sum_{j=0}^{q-1} \omega^{tr(aj)} \ket{j}\bra{j}$, where $q=p^r$ with $p$ as any prime number, $a \in \mathbb{F}_q$ and $\omega = \exp(2\pi i /p)$. In the decoding stage, the user applies projective value measurement (PVM) defined on the quotient space of the stabilizer group $\mathcal{L}(\mathcal{V})$ defined by
\begin{align}
    \mathcal{L}(\mathcal{V}) = \{c_{v} \Tilde{W}(v) : v \in \mathcal{V} \},
\end{align}
where $\mathcal{V}$ is a self orthogonal subspace in $\mathbb{F}_q ^{2N}$, 
\begin{align}
     \Tilde{W}(v) = \mathsf{X}(v_1) \mathsf{Z}(v_{N+1}) \otimes \ldots \otimes \mathsf{X}(v_N) \mathsf{Z}(v_{2N}),
\end{align}
and $c_{v} \in \mathbb{C}$ is chosen such that $\mathcal{L}(\mathcal{V})$ is an Abelian subgroup of $HW_{q}^N$  with $c_{v}I_{q^N}$ being an element of the stabilizer group if $c_{v}=1$, where $HW_{q}^N$ is the Heisenberg-Weyl group defined as,
\begin{align}
    HW_{q}^N = \{ c \Tilde{W}(s) : s \in \mathbb{F}_q^{2N}, c \in \mathbb{C} \setminus \{0\} \}.
\end{align}
In the next section, we state our main results for this problem, both for the classical and the quantum variations.

\section{Main Results}
\begin{theorem}
    For classical $X$-secure, $E$-eavesdropped, $T$-colluding SPIR (XSETSPIR) with $N$ databases, the rate given by
    \begin{align}
        R = 1 - \frac{X+\max(T,E)}{N},
    \end{align}
    is achievable, using modified cross subspace alignment (CSA) with message length $L = N-\max(T,E) - X$.
\end{theorem}
\begin{remark}
    When $X=0$ and $E=0$, the proposed scheme achieves the optimal rate for $T$-colluding SPIR, $R = 1 - \frac{T}{N}$, found in \cite{c_spir, tspir_mdscoded}.
\end{remark}
\begin{remark}
    When $X=0$, the proposed scheme achieves the optimal rate for $E$-eavesdropped, $T$-colluding SPIR,   $R = 1 - \frac{\max(T,E)}{N}$, found in \cite{C_SETPIR}.
\end{remark}
\begin{remark}
     For $X\geq1$ the exact capacity of $X$-secure PIR with a fixed number of messages $K$ is still an open problem. 
\end{remark}
\begin{theorem}
    For $X$-secure, $E$-eavesdropped, $T$-colluding quantum SPIR (XSETQSPIR) with $N$ databases which are allowed to share entanglement and have quantum channels for answer strings, the rate given by
    \begin{align}
        R_Q= \min\left\{1,2\left(1-\frac{X+\max(T,E)}{N}\right)\right\},
    \end{align}
    is achievable with modified quantum CSA.
\end{theorem}
\begin{remark}
    When $X=0$ and $E=0$, the proposed scheme achieves the capacity of $T$-colluding QSPIR, $R_Q=\min\left\{1,2\left(1-\frac{T}{N}\right)\right\}$, found in \cite{qtpir}.
\end{remark}
\section{Achievable Scheme}
Before describing the achievable scheme for the quantum $X$-secure, $E$-eavesdropped, $T$-colluding SPIR, we first introduce the classical scheme which uses the modified classical CSA to solve the classical version of the problem.

\subsection{Achievable Scheme in the Classical Setting: XSETSPIR}\label{classical scheme}
Consider a total of $N$ databases with the $T$-colluding, $E$-eavesdropped and $X$-secure setting. Let the message length $L$ be $L=N-X-M$, where $M=\max(E,T)$. The storage at each database $n$ denoted by $S_n$ is,
\begin{align}\label{classical_storage}
    S_n=\begin{bmatrix}
        W_{\cdot,1} + \sum_{i=1}^X(f_1-\alpha_n)^i R_{1i}\\
        W_{\cdot,2} + \sum_{i=1}^X(f_2-\alpha_n)^i R_{2i}\\
        \vdots\\
        W_{\cdot,L} + \sum_{i=1}^X(f_L-\alpha_n)^i R_{Li}\\
    \end{bmatrix},
\end{align}
where $W_{\cdot,j}=[W_{1,j},\ldots,W_{K,j}]^T$ is a vector representing the $j$th bit of all $K$ messages, with $W_{i,j}$ being the $j$th bit of message $i$, $R_{ij}$ are uniform independent random vectors with the same dimensions as $W_{\cdot,j}$, and $\{f_i\}_{i=1}^L$, $\{\alpha_n\}_{n=1}^N$ are globally known distinct constants from $\mathbb{F}_q$.

The user wishes to retrieve $W_{\theta}$ while protecting its privacy from any $T$ colluding databases and $E$ eavesdroppers. The user sends the query $Q_n^{[\theta]}$ to the $n$th database as,
\begin{align}\label{classical_queries}
    Q_n^{[\theta]}=\begin{bmatrix}
        \frac{\prod_{i=1}^L(f_i-\alpha_n)}{f_1-\alpha_n}\left(e_{\theta}+\sum_{i=1}^M(f_1-\alpha_n)^iZ_{1i}\right)\\
        \vdots\\
        \frac{\prod_{i=1}^L(f_i-\alpha_n)}{f_L-\alpha_n}\left(e_{\theta}+\sum_{i=1}^M(f_L-\alpha_n)^iZ_{Li} \right)
    \end{bmatrix},
\end{align}
where $e_{\theta}$ is a vector of length $K$ with a $1$ in the $\theta$th index and zero otherwise, and $Z_{ij}$ are uniform independent random vectors of length $K$ each, chosen by the user.

Since the databases want to hide any information about the messages other than the user-required message, they agree on $X+M-1$ independent uniform random variables $Z'_{1},\ldots,Z'_{X+M-1}$ before the retrieval process starts, i.e., they share common randomness, where all $X+M-1$ common randomness variables $Z'_{i}$ are random noise symbols from $\mathbb{F}_q$. Each database $n$, $n\in[1:N]$, then computes the answer to be sent to the user as,
\begin{align}\label{answers}
    A_n^{[\theta]}&=S_n^tQ_n^{[\theta]}+P_n\\
    &=\gamma_n\left(\sum_{i=1}^L\frac{1}{f_i-\alpha_n}W_{\theta,i}+\sum_{i=0}^{X+M-1}\alpha_n^i(I_i+Z'_i)\right),\label{answer2}
\end{align}
where $\gamma_n =\prod_{i=1}^L(f_i-\alpha_n)$, $P_n=\sum_{i=0}^{X+M-1}\alpha_n^iZ'_i$, and $I_i$ is the coefficient of $\alpha_n^i$ in the polynomial resulting from the product $S_n^tQ_n^{[\theta]}$. After receiving all the answer strings from the $N$ databases, the user has the following answer vector, from which the required $L$ symbols of $W_\theta$ can be obtained, as $X+M+L=N$,
\begin{align}
    &A^{[\theta]}\nonumber\\
    &= [A_1^{[\theta]},\ldots, A_N^{[\theta]}]^t \\ 
    &=B_N(\alpha,f)\nonumber\\ 
    &\quad\times[W_{\theta,1},\ldots,W_{\theta,L},I_0+Z'_0,\ldots,I_{X+M-1}+Z'_{X+M-1}]^t,
\end{align}
where $t$ represents the transpose operation, $\alpha=[\alpha_1,\ldots,\alpha_N]^t$, $f=[f_1,\ldots,f_N]^t$, and $B_N(\alpha,f)$ is an $N\times N$ invertible matrix given by, 
\begin{align}
    & B_N(\alpha,f) = \nonumber\\
    & \diag(\gamma)\begin{bmatrix}
        \!\frac{1}{f_1-\alpha_1}\!&\!\ldots\!&\!\frac{1}{f_L-\alpha_1}\!&1\!&\!\alpha_1\!&\!\ldots\!&\!\alpha_1^{X+M-1}\\
        \!\frac{1}{f_1-\alpha_2}\!&\!\ldots\!&\!\frac{1}{f_L-\alpha_2}\!&\!1\!&\!\alpha_2\!&\!\ldots\!&\!\alpha_2^{X+M-1}\\
        \!\vdots\!&\!\vdots\!&\!\vdots\!&\!\vdots\!&\!\vdots\!&\!\vdots\!&\!\vdots\\
        \!\frac{1}{f_1-\alpha_N}\!&\!\ldots\!&\!\frac{1}{f_L-\alpha_N}\!&\!1\!&\!\alpha_N\!&\!\ldots\!&\!\alpha_N^{X+M-1}
    \end{bmatrix},
\end{align}
where $\gamma=[\gamma_1,\ldots,\gamma_N]^t$. The main difference between the $N-L$ interference symbols here and the interference symbols in \cite{csa} is that they are contaminated with random noise unknown to the user, i.e., $Z'_i$ terms, which leak no information to the user except for the required $L$ bits.
\begin{remark}
    Compared to the CSA scheme, the proposed symmetric CSA scheme achieves the same rate with the extra benefit of symmetric privacy.
\end{remark}

\subsection{Achievable Scheme in the Quantum Setting: XSETQSPIR}\label{quantum scheme}

To develop the quantum scheme based on the $N$-sum box abstraction \cite{nsumbox}, we first recall some important definitions in \cite{nsumbox}.   

\begin{definition}[QCSA matrix]
The quantum CSA (QCSA) matrix of size $N\times N$ and elements from $\mathbb{F}_q$ designed to retrieve $2L$ symbols in the quantum PIR scheme is defined as follows
\begin{align}
    D_N(\alpha,\beta,f)[i,j] = \begin{cases} 
        \frac{\beta_i}{f_j-\alpha_i}, ~ j \leq L,\\
        \beta_i \alpha_i^{j-L-1}, ~ L <j\leq N,\\
    \end{cases}
\end{align}
where $\alpha=[\alpha_1,\ldots,\alpha_N]^t$, $\beta=[\beta_1,\ldots,\beta_N]^t$, $f=[f_1,\ldots,f_N]^t$, $\alpha_1,\ldots,\alpha_N,f_1,\ldots,f_L$ are distinct, $\beta_1,\ldots,\beta_N$ are non-zero and $L\leq \frac{N}{2}$.
\end{definition}

\begin{definition}[Dual QCSA matrices]\label{dual_def}
    The matrices $H^u_N$ and $H^v_N$ are defined as $H^u_N =D_N(\alpha,u,f)$ and $H^v_N =D_N(\alpha,v,f)$. Then, $H^u_N$ and $H^v_N$ are dual QCSA matrices if:
    \begin{enumerate}
        \item $u_1,\ldots,u_N$ are non-zero,
        \item $u_1,\ldots,u_N$ are distinct,
        \item for each $v_j,~j\in[1:N]$,
        \begin{align}
            v_j = \frac{1}{u_j} \left(\prod_{i=1 \atop i\neq j}^N(\alpha_j-\alpha_i)\right)^{-1}.
        \end{align}
    \end{enumerate}
\end{definition}
Using these definitions, we restate the $N$-sum box feasibility theorem from \cite[Thm.~6]{nsumbox}.
\begin{theorem}\label{feasile_nsum}
    For any dual QCSA matrices $H^u_N$ and $H^v_N$, there exists a feasible $N$-sum box transfer matrix $G(u,v)$ of size $N \times 2N$ given as follows,
    \begin{align}
        G(u,v) = G_N \begin{bmatrix}
            H^u_N & 0\\
            0& H^v_N
        \end{bmatrix}^{-1},
    \end{align}
    where
    \begin{align*}
        G_N = \begin{bmatrix}
            I_L&0_{L\times\nu} & 0 & 0 & 0 & 0\\
            0 & 0 & I_{\mu-L} & 0 & 0 & 0\\ 0 & 0 & 0 & I_L & 0_{L\times\mu} & 0\\
            0 & 0 & 0 & 0 & 0 & I_{\nu-L}
            \end{bmatrix},
    \end{align*}
    $\nu =\lceil N/2\rceil$, $\mu =\lfloor N/2\rfloor$, $I_L$ is the identity matrix of size $L\times L$, and $0_{A,B}$ is the all zeros matrix of size $A\times B$.
\end{theorem}

\begin{remark}
   We explain the the main concept behind Theorem~\ref{feasile_nsum} as follows: If $u$ and $v$ are chosen such that $H^u_N$ and $H^v_N$ are dual QCSA matrices, then there exists an $N$-entangled qubit shared between the $N$ databases such that the quantum channels between the databases and the user can be represented by $G(u,v)$.
\end{remark}

\begin{remark}
    A main difference between the quantum channel and the classical channel is that the decoding is done over-the-air. This implies that if the eavesdropper listens to $E$ answers, there is a possibility that it can get up to $E$ out of the $L$ symbols. This means that the eavesdropper is more powerful in the quantum variation compared to the classical variation.
\end{remark}

Now, we are ready to describe the XSETQSPIR scheme. The storage at each database is slightly modified compared to the classical case.  The storage in the quantum scheme $S_Q$ is given as, 
\begin{align}
    S_Q = [S_n(1)^t, ~S_n(2)^t]^t,
\end{align}
where $S_n(1)$ and $S_n(2)$ are as in \eqref{classical_storage}, i.e., each containing $L = N-X-M \leq \frac{N}{2}$ new symbols of the $K$ messages, along with new random noise vectors. In other words, the length of the messages considered in the quantum scheme is twice of what was considered in the classical case. To retrieve the required message, the user sends the query $Q_n^{[\theta]}$ to database $n$, which is of the same form as in the classical scheme in \eqref{classical_queries}. Each database $n$, $n\in[1:N]$, then generates the noise added answers as in \eqref{answers}, 
\begin{align}\label{quantum_sub_answers}
    \hat{A}_n^{[\theta]}(1)&=S_n(1)^tQ_n^{[\theta]}+P_n(1)\\
    \hat{A}_n^{[\theta]}(2)&=S_n(2)^tQ_n^{[\theta]}+P_n(2)
\end{align}
where 
\begin{align} 
P_n(1)=\!\!\sum_{i=0}^{X+M-1}\alpha_n^iZ'_i(1), \quad P_n(2)=\!\!\sum_{i=0}^{X+M-1}\alpha_n^iZ'_i(2)
\end{align}
with all $Z'_i(j)$ being random noise symbols. To prevent the eavesdropper from decoding over-the-air, the user sends two masking variables to each database $n$, given by,
\begin{align}\label{masking}
    &\Lambda_n(\kappa)\nonumber\\
    &=\gamma_n
        \Big(\frac{1}{f_1-\alpha_n}\lambda_{1}(\kappa)+\ldots+\frac{1}{f_L-\alpha_n}\lambda_{L}(\kappa)\nonumber\\
        &\quad+\lambda_{L+1}(\kappa)+ \alpha_n \lambda_{L+2}(\kappa)+ \ldots+\alpha_n^{N-L-1} \lambda_{N}(\kappa)\Big),
\end{align}
for $\kappa\in[1,2]$, where $\lambda_{n}(\kappa), ~n\in[1:N],~\kappa \in [1:2] $ are uniform independent random variables generated by the user. Then, each database generates two answer instances $A_n^{[\theta]}(1),~A_n^{[\theta]}(2)$,
\begin{align}
    A_n^{[\theta]}(1) &= \hat{A}_n^{[\theta]}(1)+\Lambda_n(1)\\
    A_n^{[\theta]}(2) &= \hat{A}_n^{[\theta]}(2)+\Lambda_n(2).
\end{align}
The $N$ initial answers from the $N$ databases are written compactly as,
\begin{align}
    A\!\!&=\!\![A^{[\theta]}_1(1),\ldots,A_N^{[\theta]}(1),A_1^{[\theta]}(2),\ldots,A_N^{[\theta]}(2)]^t\!\!\\&=\!\!\begin{bmatrix}
        \text{diag}(\gamma)&0\\
        0&\text{diag}(\gamma)
    \end{bmatrix}\!\!
    \begin{bmatrix}
        D_N\!\!&\!\!0\\
        0\!\!&\!\!D_N
    \end{bmatrix}\!\!\begin{bmatrix}
       X(1)\\
       X(2)
    \end{bmatrix}
\end{align}
where $D_N=D_N(\alpha,1_N,f)$, $\gamma=[\gamma_1,\ldots,\gamma_N]^t$, and 
\begin{align}\label{transmitted_symbols}
     X(i)\!=\!\begin{bmatrix}
        W_{\theta,1}(i)+\lambda_{1}(i)\\
        W_{\theta,2}(i)+\lambda_{2}(i)\\
        \vdots\\
        W_{\theta,L}(i)+\lambda_{L}(i)\\
        I_0(i)+Z'_0(i)+ \lambda_{L+1}(i)\\
        \vdots\\
        I_{X+M-1}(i)+Z'_{X+M-1}(i)+  \lambda_{N}(i)
    \end{bmatrix}.
\end{align}
for $i \in [1:2]$. Then, to make use of the entanglement and quantum channels, the answers are modified as,
\begin{align}
    \Tilde{A}=\begin{bmatrix}
        \text{diag}(u)&0\\
        0&\text{diag}(v)
    \end{bmatrix}\begin{bmatrix}
        \text{diag}(\gamma)&0\\
        0&\text{diag}(\gamma)
    \end{bmatrix}^{-1} A,
\end{align}
where $u=[u_1,\ldots,u_N]^t$ and $v=[v_1,\ldots,v_N]^t$ are chosen such that they satisfy Definition~\ref{dual_def}. These answers are sent through the quantum channels using the encoder defined by the Pauli operators, i.e., each database sends its answer instances $\Tilde{A}_n(1), \Tilde{A}_n(2)$, $n \in [1:N]$ as follows,
\begin{align}
    \mathcal{A}_n^{[\theta]} = \mathsf{Z}(\Tilde{A}_n(2))\mathsf{X}(\Tilde{A}_n(1))\mathcal{A}_n^0.
\end{align}
Based on the properties of the quantum channel, the $N$ symbols received by the user, denoted by $y$ are given as, 
\begin{align}
    y &= G(u,v) \Tilde{A}\\
    &=G(u,v)\begin{bmatrix}
        H_N^u&0\\
        0&H_N^v
    \end{bmatrix}\begin{bmatrix}
        X(1)\\
        X(2)
    \end{bmatrix}\\
    &=G_N \begin{bmatrix}
        X(1)\\
        X(2)
    \end{bmatrix}\\
    &= [W_{\theta,1}(1)+\lambda_{1}(1),\ldots,W_{\theta,L}(1)+\lambda_{L}(1),I'(1),\nonumber\\
    & \ \quad W_{\theta,1}(2)+\lambda_{1}(2),\ldots,W_{\theta,L}(2)+\lambda_{L}(2),I'(2)]^t,
    \end{align}
where $I'(1)$ represents the last $\lfloor N/2 \rfloor -L$ interference symbols of $X(1)$ in \eqref{transmitted_symbols}, and $I'(2)$ represent the last $\lceil N/2 \rceil -L$ interference symbols of $X(2)$ in \eqref{transmitted_symbols}. As the user already knows the values of $\lambda_\ell(\kappa)$ for $\ell\in[1:L]$ and $\kappa\in[2]$, the user obtains the $2L$ symbols of the required message $W_\theta$, denoted by $W_{\theta,1}(1),\dotsc,W_{\theta,L}(1),W_{\theta,1}(2),\dotsc,W_{\theta,L}(2)$.

\begin{remark}
    In this scheme, we use the fact that the length of each message sub-packet $L$ must satisfy both $L=N-X-\max(T,E)$, and $L\leq \frac{N}{2}$. If $L > \frac{N}{2}$, we drop the extra databases as in \cite{nsumbox}.
\end{remark}

\begin{remark}
    Note that since $u$ and $v$ can be globally known, the no-cloning theorem cannot be invoked, thus the eavesdropper can listen to quantum channels.
\end{remark}

\begin{remark}
    Due to the over-the-air decoding, the user needs to send masking variables, $\lambda_1,\ldots,\lambda_N$, to the $N$ databases, thus $N^2$ bits in total. However, in our proposed scheme the user needs only to send $1$ bit to each database over the non-secure channel, i.e., $N$ bits in total, to achieve the same goal. 
\end{remark}

\section{Conclusions}

In this paper, we studied the classical and quantum variations of the $X$-secure, $E$-eavesdropped, and $T$-colluding symmetric PIR. In the classical variation, we developed a scheme that achieves symmetric privacy at the same rate as the state-of-the-art scheme that solves the same problem without symmetric privacy. In the quantum variation, we uncovered how the eavesdroppers have better access to the transmitted answer strings due to the over-the-air decodability imposed by the $N$-sum box abstraction. To that end, we designed a scheme that represses over-the-air decodability while maintaining the super-dense coding gain, i.e., doubling the rate compared to the classical variation.

\bibliographystyle{unsrt}
\bibliography{references}

\begin{thebibliography}{10}

\bibitem{chor}
B.~Chor, E.~Kushilevitz, O.~Goldreich, and M.~Sudan.
\newblock Private information retrieval.
\newblock {\em Jour. of the ACM}, 45(6):965--981, November 1998.

\bibitem{c_pir}
H.~Sun and S.~A. Jafar.
\newblock The capacity of private information retrieval.
\newblock {\em IEEE Trans. Info. Theory}, 63(7):4075--4088, July 2017.

\bibitem{c_spir}
H.~Sun and S.~A. Jafar.
\newblock The capacity of symmetric private information retrieval.
\newblock {\em IEEE Trans. Info. Theory}, 65(1):322--329, June 2018.

\bibitem{mdstpir}
H.~Sun and S.~A. Jafar.
\newblock Private information retrieval from {MDS} coded data with colluding
  servers: Settling a conjecture by {F}reij-{H}ollanti et al.
\newblock {\em IEEE Trans. Info. Theory}, 64(2):1000--1022, December 2017.

\bibitem{tspir_mdscoded}
Q.~Wang and M.~Skoglund.
\newblock Symmetric private information retrieval from {MDS} coded distributed
  storage with non-colluding and colluding servers.
\newblock {\em IEEE Trans. Info. Theory}, 65(8):5160--5175, March 2019.

\bibitem{C_SETPIR}
Q.~Wang, H.~Sun, and M.~Skoglund.
\newblock The capacity of private information retrieval with eavesdroppers.
\newblock {\em IEEE Trans. Info. Theory}, 65(5):3198--3214, December 2018.

\bibitem{first_xsecure}
H.~Yang, W.~Shin, and J.~Lee.
\newblock Private information retrieval for secure distributed storage systems.
\newblock {\em IEEE Trans. Info. Foren. Security}, 13(12):2953--2964, May 2018.

\bibitem{csa}
Z.~Jia, H.~Sun, and S.~A. Jafar.
\newblock Cross subspace alignment and the asymptotic capacity of {$X$}-secure
  {$T$}-private information retrieval.
\newblock {\em IEEE Trans. Info. Theory}, 65(9):5783--5798, May 2019.

\bibitem{arbitrarycollusion}
X.~Yao, N.~Liu, and W.~Kang.
\newblock The capacity of private information retrieval under arbitrary
  collusion patterns for replicated databases.
\newblock {\em IEEE Trans. Info. Theory}, 67(10):6841--6855, July 2021.

\bibitem{byzantine_tpir}
K.~Banawan and S.~Ulukus.
\newblock The capacity of private information retrieval from {B}yzantine and
  colluding databases.
\newblock {\em IEEE Trans. Info. Theory}, 65(2):1206--1219, September 2018.

\bibitem{banawan_eaves}
K.~Banawan and S.~Ulukus.
\newblock Private information retrieval through wiretap channel {II}: Privacy
  meets security.
\newblock {\em IEEE Trans. Info. Theory}, 66(7):4129--4149, February 2020.

\bibitem{banawan_multimessage_pir}
K.~Banawan and S.~Ulukus.
\newblock Multi-message private information retrieval: Capacity results and
  near-optimal schemes.
\newblock {\em IEEE Trans. Info. Theory}, 64(10):6842--6862, April 2018.

\bibitem{banawan_pir_mdscoded}
K.~Banawan and S.~Ulukus.
\newblock The capacity of private information retrieval from coded databases.
\newblock {\em IEEE Trans. Info. Theory}, 64(3):1945--1956, January 2018.

\bibitem{uncoded_constrainedstorage_pir}
M.~A. Attia, D.~Kumar, and R.~Tandon.
\newblock The capacity of private information retrieval from uncoded storage
  constrained databases.
\newblock {\em IEEE Trans. Info. Theory}, 66(11):6617--6634, September 2020.

\bibitem{batuhan_hetero}
K.~Banawan, B.~Arasli, Y.-P. Wei, and S.~Ulukus.
\newblock The capacity of private information retrieval from heterogeneous
  uncoded caching databases.
\newblock {\em IEEE Trans. Info. Theory}, 66(6):3407--3416, June 2020.

\bibitem{grpahbased_pir}
N.~Raviv, I.~Tamo, and E.~Yaakobi.
\newblock Private information retrieval in graph-based replication systems.
\newblock {\em IEEE Trans. Info. Theory}, 66(6):3590--3602, November 2019.

\bibitem{ITW_paper}
K.~Banawan, B.~Arasli, and S.~Ulukus.
\newblock Improved storage for efficient private information retrieval.
\newblock In {\em IEEE ITW}, August 2019.

\bibitem{multimessage_pir_sideinfo}
M.~J. Siavoshani, S.~P. Shariatpanahi, and M.~Ali Maddah-Ali.
\newblock Private information retrieval for a multi-message scenario with
  private side information.
\newblock {\em IEEE Trans. on Commun.}, 69(5):3235--3244, January 2021.

\bibitem{wei_banawan_cache_pir}
Y.-P. Wei, K.~Banawan, and S.~Ulukus.
\newblock Fundamental limits of cache-aided private information retrieval with
  unknown and uncoded prefetching.
\newblock {\em IEEE Trans. Info. Theory}, 65(5):3215--3232, November 2018.

\bibitem{wang_spir}
Z.~Wang and S.~Ulukus.
\newblock Symmetric private information retrieval at the private information
  retrieval rate.
\newblock {\em IEEE Jour. on Selected Areas in Info. Theory}, 3(2):350--361,
  June 2022.

\bibitem{semantic_pir}
S.~Vithana, K.~Banawan, and S.~Ulukus.
\newblock Semantic private information retrieval.
\newblock {\em IEEE Trans. Info. Theory}, 68(4):2635--2652, December 2021.

\bibitem{salim_singleserver_pir}
A.~Heidarzadeh, S.~Kadhe, S.~El Rouayheb, and A.~Sprintson.
\newblock Single-server multi-message individually-private information
  retrieval with side information.
\newblock In {\em IEEE ISIT}, July 2019.

\bibitem{Salim_CodedPIR}
R.~Tajeddine, O.~Gnilke, and S.~El Rouayheb.
\newblock Private information retrieval from {MDS} coded data in distributed
  storage systems.
\newblock {\em IEEE Trans. Info. Theory}, 64(11):7081--7093, March 2018.

\bibitem{recent}
S.~Ulukus, S.~Avestimehr, M.~Gastpar, S.~A. Jafar, R.~Tandon, and C.~Tian.
\newblock Private retrieval, computing and learning: Recent progress and future
  challenges.
\newblock {\em IEEE Jour. on Selected Areas in Commun.}, 40(3):729--748, March
  2022.

\bibitem{qpir}
S.~Song and M.~Hayashi.
\newblock Capacity of quantum private information retrieval with multiple
  servers.
\newblock {\em IEEE Trans. Info. Theory}, 67(1):452--463, September 2021.

\bibitem{qtpir}
S.~Song and M.~Hayashi.
\newblock Capacity of quantum private information retrieval with colluding
  servers.
\newblock {\em IEEE Trans. Info. Theory}, 67(8):5491--5508, May 2021.

\bibitem{qtpir_t=n-1}
S.~Song and M.~Hayashi.
\newblock Capacity of quantum symmetric private information retrieval with
  collusion of all but one of servers.
\newblock {\em IEEE Jour. on Selected Areas in Info. Theory}, 2(1):380--390,
  January 2021.

\bibitem{qpir_colluding_mdscoded}
M.~Allaix, S.~Song, L.~Holzbaur, T.~Pllaha, M.~Hayashi, and C.~Hollanti.
\newblock On the capacity of quantum private information retrieval from
  mds-coded and colluding servers.
\newblock {\em IEEE Jour. on Selected Areas in Commun.}, 40(3):885--898,
  January 2022.

\bibitem{noisy_qpir}
Y.~Yang, P.~Yang, G.~Xu, Y.~Zhou, and W.~Shi.
\newblock Quantum private information retrieval over a collective noisy
  channel.
\newblock {\em Modern Physics Letters A}, 38(01):2350001, 2023.

\bibitem{qmqpir}
S.~Song and M.~Hayashi.
\newblock Quantum private information retrieval for quantum messages.
\newblock In {\em IEEE ISIT}, September 2021.

\bibitem{qpir_byzantine_colluding}
P.~Saarela, M.~Allaix, R.~Freij-Hollanti, and C.~Hollanti.
\newblock Private information retrieval from colluding and byzantine servers
  with binary reed–muller codes.
\newblock In {\em IEEE ISIT}, August 2022.

\bibitem{qpir_star_product_codes}
M.~Allaix, L.~Holzbaur, T.~Pllaha, and C.~Hollanti.
\newblock High-rate quantum private information retrieval with weakly self-dual
  star product codes.
\newblock In {\em IEEE ISIT}, July 2021.

\bibitem{nsumbox}
M.~Allaix, Y.~Lu, Y.~Yao, T.~Pllaha, C.~Hollanti, and S.~Jafar.
\newblock {$N$}-sum box: An abstraction for linear computation over many-to-one
  quantum networks.
\newblock 2023.
\newblock Available online at arXiv:2304.07561.

\bibitem{nielsen-chuang}
M.~Nielsen and I.~Chuang.
\newblock {\em Quantum Computation and Quantum Information: 10th Anniversary
  Edition}.
\newblock Cambridge University Press, 2010.

\end{thebibliography}
\end{document}